\documentclass[acmsmall,screen,authordraft,nonacm,review=false,timestamp=false]{acmart}
\AtBeginDocument{%
  }

\setcopyright{acmlicensed}
\copyrightyear{2025}
\acmYear{2025}
\acmDOI{XXXXXXX.XXXXXXX}
\acmConference[CSCW Companion '25]{Companion Publication of the 2025 Conference on Computer-Supported Cooperative Work and Social Computing}{Oct 18--22,
  2025}{Woodstock, NY}
\acmISBN{978-1-4503-XXXX-X/2018/06}

\begin{document}

\title[Virtual Interviewers, Real Results: Exploring AI-Driven Mock Technical Interviews]{Virtual Interviewers, Real Results: Exploring AI-Driven Mock Technical Interviews on Student Readiness and Confidence}

\author{Nathalia Gomez}
\authornote{Both authors contributed equally to this research.}
\email{nmg88@drexel.edu}
\author{S. Sue Batham}
\authornotemark[1]
\email{sb4529@drexel.edu}
\affiliation{%
  \institution{Drexel University}
  \city{Philadelphia}
  \state{Pennsylvania}
  \country{USA}
}

\author{Matias Volonte}
\affiliation{%
  \institution{Clemson University}
  \city{Clemson}
  \state{South Carolina}
  \country{USA}
}
\email{mvolont@clemson.edu}

\author{Tiffany D. Do}
\affiliation{%
  \institution{Drexel University}
  \city{Philadelphia}
  \state{Pennsylvania}
  \country{USA}
}
\email{tiffany.do@drexel.edu}

\renewcommand{\shortauthors}{Gomez and Batham et al.}

\begin{abstract}
Technical interviews are a critical yet stressful step in the hiring process for computer science graduates, often hindered by limited access to practice opportunities. This formative qualitative study (n=20) explores whether a multimodal AI system can realistically simulate technical interviews and support confidence-building among candidates. Participants engaged with an AI-driven mock interview tool featuring whiteboarding tasks and real-time feedback. Many described the experience as realistic and helpful, noting increased confidence and improved articulation of problem-solving decisions. However, challenges with conversational flow and timing were noted. These findings demonstrate the potential of AI-driven technical interviews as scalable and realistic preparation tools, suggesting that future research could explore variations in interviewer behavior and their potential effects on candidate preparation.
\end{abstract}




\maketitle

\section{Introduction}
Technical interviews are a critical part of the hiring process for computer science graduates, with major companies like Google, Amazon, and Meta requiring them \cite{behroozi2019, Cui2024}. Despite their prevalence, these interviews often induce significant stress, leaving many students feeling unprepared and lacking confidence. Lunn et al. \cite{lunn2021} found that feeling underprepared is one of the most common negative experiences tied to technical interviews. Research also suggests that disparities in access to preparation resources can impact students' readiness, as the most common way to practice is mock interviews with peers \cite{lunnUneven2021}. However, students with fewer connections in computing can struggle to find mock interview partners, and those with outside obligations, such as part-time jobs or caregiving responsibilities, have limited time to engage in mock sessions, thus exacerbating inequities in the preparation process \cite{lunnUneven2021}.

Advances in generative AI have enabled the development of virtual interviewers that can simulate aspects of technical interviews in real time, analyzing multiple modalities such as voice and pseudocode \cite{gpt4o}. To explore the potential of AI to support students in preparing for technical interviews, we conducted a formative user study with a preliminary AI-driven mock interview system. This study examines the feasibility of using AI to assist with interview preparation, with a focus on how such systems might influence student confidence and perceived readiness. If AI systems can approximate key elements of real interviews, their generative flexibility could offer opportunities to vary interviewer behaviors, such as tone, feedback style, and visual presence, to investigate how these factors shape candidate experiences. This formative work aims to inform future efforts in refining AI-driven tools by addressing the following research questions: (1) How effectively can an AI-driven mock interview system simulate technical interviews? (2) How does practicing with an AI interviewer impact students' confidence, perceived readiness, and ability to articulate engineering decisions? (3) What are the advantages and limitations of AI-driven mock interviews compared to traditional preparation methods?

\section{Related Work}
\subsection{Challenges in Technical Interviews}
Technical interviews, defined as hiring interviews for computing positions that occur online, via phone/video call, or on-site/in-person, and include any combination of problem solving, coding, or programming tests for job candidates” \cite{lunnUneven2021}, are a crucial step in securing computing jobs. These often involve whiteboarding tasks where candidates solve problems while explaining their thought process \cite{behroozi2020, kapoor2020}. While companies expect similar problem-solving skills across roles \cite{Ford2017}, many candidates struggle with preparation, often needing months to feel ready \cite{lunnUneven2021}. Despite this effort, they frequently experience self-doubt, which can lower self-efficacy and discourage applications for internships and jobs \cite{kapoor2020}. Low self-efficacy hinders performance under pressure, even for technically proficient candidates \cite{halljrEffects2018}. Anxiety can further erode confidence \cite{behrooziDoes2020}, and many lack experience verbalizing their thought process in real time. Our AI-driven simulation addresses these challenges by providing structured practice, helping candidates refine technical skills and articulate engineering decisions to improve self-efficacy and interview readiness.

\subsection{Technical Interview Simulations}
Past efforts to simulate technical interviews have been limited by the capabilities of existing technologies. For instance, Salvi et al. \cite{Salvi2017} developed a system that used semantic analysis and Google Cloud Speech-to-Text to assess candidate responses. However, the system experienced latency that disrupted conversational flow and lacked multimodal capabilities needed to support whiteboarding or real-time code analysis. Similarly, Chou et al. \cite{Chou2022} designed an AI-guided interview platform for general interview practice. While their system used pose estimation and feature tracking to monitor nonverbal behavior, it did not include technical problem-solving tasks or interactive virtual agents. These limitations restricted their usefulness for simulating realistic technical interviews. In contrast, our system leverages recent advances in generative AI to support real-time dialogue, code evaluation, and multimodal interaction, offering a more complete simulation of the technical interview experience.

\section{AI-Driven Mock Technical Interview System}

For our study, we developed a real-time, AI-driven agent that replicates the structure and dynamics of technical interviews by presenting questions, engaging in dialogue, and providing feedback. The system combines a Unity frontend with a backend powered by the open-source LiveKit toolkit, supporting dynamic interaction and future integration with virtual avatars. The system uses GPT-4o \cite{openai2025chatgpt} for multimodal language processing, Deepgram for speech transcription, and Silero VAD \cite{SileroVAD} to detect when users were speaking. A fine-tuned SmolLM v2 model \cite{allal2025smollm2smolgoesbig} predicts speech boundaries to support smooth turn-taking, while spoken responses are generated using GPT-4o mini TTS with the Onyx voice. When a user initiates a call in Unity, the system launches an agent and virtual meeting room via LiveKit. The Unity interface included a basic code textbox with syntax highlighting and auto-indentation, without codesense or autocomplete, similar to traditional whiteboard interview platforms. The system captures the code editor content every three seconds to monitor and analyze user progress in real time. With an average response latency of ~300ms—comparable to human conversation \cite{Stivers2009}—our prototype demonstrates the feasibility of AI-driven technical interview simulations that provide interactive, low-latency feedback on both code and verbal explanations. This extensible architecture could possibly support future research into interviewer behaviors, nonverbal cues, and personalized interview dynamics.

\section{Methods}
The study was a 50-minute in-person session approved by an Institutional Review Board. We recruited 20 participants (12 men, 8 women) from a U.S. university’s College of Computing and Informatics using mailing lists and flyers. All were juniors, seniors, or graduate students actively preparing for technical interviews. Participants first completed an online screener capturing demographics, including ethnicity, education, and prior interview experience. Afterwards, they were briefed on the study and informed they would be completing a technical interview with an AI agent. To support comfort and privacy, participants wore headphones and completed a whiteboarding-style interview using a “Medium”-level LeetCode problem, which none had previously encountered\footnote{\url{https://leetcode.com/problems/h-index/description/}}. All participants received the same question to ensure consistency. Following the mock interview, participants completed a semi-structured interview reflecting on realism, overall experience, and perceived impact on confidence. Detailed demographics, recorded sample mock interview, and interview questions are included in our supplemental materials: \url{https://doi.org/10.17605/OSF.IO/627TS}.

\section{Results}
A qualitative analysis was conducted using inductive techniques to extract insights from participants' experiences with the mock interview system. We conducted a grounded thematic analysis following Braun and Clarke’s six-phase framework, systematically reviewing and coding semi-structured interview transcripts to identify key themes. Using template coding, we categorized responses based on relevance (e.g., difficulty customization, confidence boost, interview skills improvement) and refined them iteratively. The first and second authors collaboratively coded the data, reaching consensus through discussion, while the remaining authors provided guidance in interpreting the findings, ensuring rigor in line with McDonald’s recommendations for CSCW and HCI research \cite{mcdonald_reliability_2019}.

\subsection {AI Interaction Felt Natural and Effective}

The participants’ experiences revealed that the interaction between them and the AI felt human-like, natural, and effective for a technical interviewer. Most participants (80\%, N = 16) found the AI’s speech and conversational style to be realistic, often expressing surprise at the natural and human-like quality of its voice. P17 noted, "\textit{I think it felt very human, like it used some filler words that made me feel like I was almost talking to another human being. The conversation was like just talking to a phone agent.}”

Beyond its speech patterns, participants also observed that the AI conducted itself in a manner similar to real technical interviewers. Several participants highlighted how the AI followed structured interview protocols, such as explaining concepts, providing coding examples, and giving hints when needed. P9 remarked, "\textit{It’s very similar to what an actual employer would do during technical interviews}," while P10 emphasized, "\textit{It explained the concepts to me like a real software engineer would—that was really good.}" Participants also appreciated the AI’s ability to guide them through the problem-solving process, with P17 stating, "\textit{The realistic parts were definitely how it made you go through examples and [asked] you how to work around them.}” Moreover, participants who had previously experienced human-led mock interviews found the AI's behavior strikingly familiar when it came to common interviewing practices, with P8 stating, "\textit{It asked me to run through test cases at the end, which is something I’ve seen in real mock interviews.}"

\subsection{Interview Skills Improvement}

Participants also identified key interview skills that they felt were strengthened by engaging in the AI-led mock interview. One of the most frequently mentioned improvements was in articulating one’s thought process. Participants emphasized that effectively explaining problem-solving approaches is a skill often valued by real interviewers. As P08 observed, "\textit{Interviewers are often complaining about this sort of thing where the candidates can solve the problems within 20 minutes, but they don't talk, or they don't communicate. So, I feel like this will help a ton with the communication aspect.}"

Another interesting insight that emerged was the system’s ability to simulate the pressure of being observed while problem-solving. Participants described this experience in different ways. For example, P11 referred to it as “\textit{being on the spot” or “having someone watching over you.}” P13 explained, “\textit{When you're doing problems on your own, [...] you feel like you have a lot of time. And you know there isn't that pressure of, you know, being on a time limit of having someone seeing you or monitoring you, that you're gonna get in a real interview.}" P11 further reinforced this point, stating, “\textit{There's this added intensity which I thought was awesome for interview prep.}” Together, these reflections suggest that the AI system not only targets a well-known communication gap but also provides a realistic and high-pressure environment that better mirrors the psychological conditions of real interviews.

\subsection{Perceived Usefulness of AI as a Realistic Interview Preparation Tool}
Nearly all of our participants (80\%, N = 16) found the AI-driven mock interview system to be a useful and beneficial tool for technical interview preparation. In fact, 65\% (N = 13) of our participants said they would use this system again, citing its ability to simulate real interview conditions while providing structured guidance and feedback. For example, P17 highlighted the value of the conversational interaction, saying, \textit{"I definitely can see how this can be very helpful, especially the speech of the interaction"}. Nearly half of our participants (40\%, N = 8) also explicitly highlighted how the system compared favorably to existing preparation methods. 

Many felt that AI-led mock interviews offered advantages over platforms like LeetCode (P1, P2, P11, P20), where feedback is limited to hints rather than interactive guidance. Others showed preference over the system's ability to encourage candidates to articulate their thought process, an essential skill often overlooked in other interview prep platforms. As P14 explained, “\textit{The environment is very much like the technical interviews where you are speaking to somebody explaining your thought process and writing code at the same time, regardless of what the other tools are and can provide for you. They don't do that experience.}"

\subsection{Confidence Boost}
Participants frequently described the AI mock interview system as a confidence-boosting tool, thanks to its low-stakes, judgment-free environment. This setup allowed many to engage more openly with the task, with 60\% (N = 12) noting that the lack of human evaluation reduced anxiety and made it easier to focus. As P15 participant shared, \textit{"For when you're practicing with peers, sometimes you feel stressed about what they will think of you... [this is] not a real person, and you don't feel that stressed about it"}. Others described the experience as a realistic simulation of a technical interview, but without the stress of being judged. \textit{"It reminded me of what a real interview is like, but less scary. I’d feel more comfortable going into a real one now"} -- P13.

Additionally, some participants described the potential for repeated use of the mock interview system as resembling exposure therapy. As P18 participant noted, \textit{"I think if I used this a few more times, I’d feel a lot better walking into real interviews. It’s like exposure therapy in a way"}. These reflections suggest the system’s potential to reduce interview anxiety and improve preparedness, offering a realistic yet forgiving environment to build confidence for real-world opportunities.

\subsection{Perceived Issues with AI Conversational Flow \& Responsiveness}
Participants were also asked to identify aspects of the system that felt unrealistic. The most commonly cited issues (85\%, N=17) related to the AI’s conversational style and response timing. Some participants noted that slow response times disrupted the flow of the interview. As P18 summarized concisely, “\textit{Maybe just the time responding wasn't as accurate as a person would be.}” Others, conversely, found the AI’s speech speed to be too fast at times, making it difficult to follow. P08 described this experience, stating, “\textit{At times I feel like it was speaking a little bit too fast for me to understand.}” 

These findings reveal that participants were highly attuned to the AI’s conversational rhythm and flow. Since a key goal of this system is to provide an interview experience that effectively prepares students, ensuring a natural, well-paced interaction is crucial. If the AI’s delivery disrupts comprehension or engagement, it may hinder rather than help students build confidence and readiness. Addressing these conversational nuances could further enhance the realism of AI-driven mock interviews, making them an even more effective tool for technical interview preparation.

\subsection{Desire for Visual and Interactive Features}
The mock interview system was designed with a simplified interface, but many participants felt the experience lacked key elements needed to fully engage in technical problem-solving. More than half of participants (55\%, N = 11) expressed a desire for more visual cues and interactive features to support their understanding during the session. A common suggestion was the ability to run code or see test cases, which participants saw as essential for debugging and understanding the task. This was an interesting observation, since these participants also noted that real whiteboarding interviews do not include these features. These requested features, such as the ability to compile code, were seen as useful for practice rather than direct interview simulations, suggesting the need for a flexible system that supports both preparation and realistic interview conditions.  

\subsection{Personalization Options}
Participants reflected on how the AI mock interview system could better accommodate individual preferences and preparation needs. About 45\% of participants (N = 9) emphasized the importance of adaptable and customizable features. Several suggested the option to select a difficulty level---beginner, intermediate, or advanced---before starting the session. This would allow the interviewee to feel more appropriately matched to their skill level. As P15 shared, \textit{"The interviewer assumed I’m like a beginner... it probably needs to be personalized more, like you can choose the difficulty"}. Others expressed interest in controlling the level of support provided, with some preferring guidance only when requested to challenge themselves, while others desired a more forgiving mode for practice sessions. Some participants highlighted the importance of setting expectations before the session. A few also proposed customization features to enhance comfort, such as choosing between different voice styles. These insights suggest that adding more flexible and personalized options could help make the experience feel more natural, user-centered, and supportive of individual goals.

\section{Discussion}
\subsection{AI Can Accurately Mimic Technical Interviews}
The combined experiences of our participants strongly indicate that AI-led mock interviews can effectively replicate the conditions of real technical interviews (RQ1). Many participants explicitly highlighted key elements of the system that contributed to this realism, including the sense of being monitored, time pressure, real-time guidance and interventions, and the encouragement to articulate their thought process. These features set AI-driven mock interviews apart from traditional, asynchronous preparation methods, which lack the interactive and dynamic nature of a live interview. 

Despite minor conversational flow issues noted by some participants, the overall perception was that the AI’s ability to engage in real-time dialogue and adapt to responses was a major advantage. More than half of our participants found the AI’s human-like interactions to be highly realistic, reinforcing the system’s effectiveness in preparing candidates for actual technical interviews. Notably, interview experiences vary widely. Some interviewers write out problems, while others rely on verbal explanations; some offer frequent hints, whereas others remain passive observers. Given this variability, AI-led interviews can possibly expose candidates to a broader range of possible interview dynamics, making them highly adaptable. 

\subsection{AI Technical Interviews Increase Confidence and Help Preparation}

Across interviews, participants shared how the AI mock interview system contributed to their confidence and readiness for real-world technical interviews (RQ2/RQ3). Many described the experience as both realistic and approachable, noting that it reduced anxiety and allowed them to mentally rehearse the structure of a technical interview without feeling judged. Participants consistently emphasized the value of practicing in a safe, low-pressure setting, which enabled them to focus on their thought process rather than performance and made them feel more capable of navigating future interviews. Some also viewed the system as a valuable long-term tool, with repeated use acting as a form of exposure therapy to gradually build resilience and a growth mindset. These insights highlight the potential of AI mock interviews to build confidence and reduce interview-related anxiety, particularly for students with limited access to human interview partners or formal coaching.

\subsection{Future Work}

Our study offers a preliminary exploration of how AI-driven mock interviews can affect candidate preparation for technical interviews. Participants found the experience both realistic and confidence-building, suggesting that future work should investigate how variations in interviewer behavior—such as tone, responsiveness, and feedback style—impact student confidence, articulation, and performance under pressure. Additionally, we plan to incorporate 3D avatars to explore how nonverbal cues and visual presence influence candidate confidence and behavior. These investigations could offer valuable insights for enhancing interviewer training practices in the industry.

Building on participants’ interest in more tailored experiences, future work should also explore personalization features like adjustable difficulty, guidance levels, and interviewer demeanor. These enhancements may help meet individual needs, reduce anxiety, and support skill development. Finally, future studies should also examine how repeated use supports long-term preparation and confidence over time.


\bibliographystyle{ACM-Reference-Format}
\bibliography{sample-base}


\begin{thebibliography}{17}


\ifx \showCODEN    \undefined \def \showCODEN     #1{\unskip}     \fi
\ifx \showISBNx    \undefined \def \showISBNx     #1{\unskip}     \fi
\ifx \showISBNxiii \undefined \def \showISBNxiii  #1{\unskip}     \fi
\ifx \showISSN     \undefined \def \showISSN      #1{\unskip}     \fi
\ifx \showLCCN     \undefined \def \showLCCN      #1{\unskip}     \fi
\ifx \shownote     \undefined \def \shownote      #1{#1}          \fi
\ifx \showarticletitle \undefined \def \showarticletitle #1{#1}   \fi
\ifx \showURL      \undefined \def \showURL       {\relax}        \fi
\providecommand\bibfield[2]{#2}
\providecommand\bibinfo[2]{#2}
\providecommand\natexlab[1]{#1}
\providecommand\showeprint[2][]{arXiv:#2}

\bibitem[gpt(2024)]%
        {gpt4o}
 \bibinfo{year}{2024}\natexlab{}.
\newblock \bibinfo{title}{Hello GPT-4o}.
\newblock
\urldef\tempurl%
\url{https://openai.com/index/hello-gpt-4o/}
\showURL{%
\tempurl}


\bibitem[Allal et~al\mbox{.}(2025)]%
        {allal2025smollm2smolgoesbig}
\bibfield{author}{\bibinfo{person}{Loubna~Ben Allal}, \bibinfo{person}{Anton Lozhkov}, \bibinfo{person}{Elie Bakouch}, \bibinfo{person}{Gabriel~Martín Blázquez}, \bibinfo{person}{Guilherme Penedo}, \bibinfo{person}{Lewis Tunstall}, \bibinfo{person}{Andrés Marafioti}, \bibinfo{person}{Hynek Kydlíček}, \bibinfo{person}{Agustín~Piqueres Lajarín}, \bibinfo{person}{Vaibhav Srivastav}, \bibinfo{person}{Joshua Lochner}, \bibinfo{person}{Caleb Fahlgren}, \bibinfo{person}{Xuan-Son Nguyen}, \bibinfo{person}{Clémentine Fourrier}, \bibinfo{person}{Ben Burtenshaw}, \bibinfo{person}{Hugo Larcher}, \bibinfo{person}{Haojun Zhao}, \bibinfo{person}{Cyril Zakka}, \bibinfo{person}{Mathieu Morlon}, \bibinfo{person}{Colin Raffel}, \bibinfo{person}{Leandro von Werra}, {and} \bibinfo{person}{Thomas Wolf}.} \bibinfo{year}{2025}\natexlab{}.
\newblock \bibinfo{title}{SmolLM2: When Smol Goes Big -- Data-Centric Training of a Small Language Model}.
\newblock
\showeprint[arxiv]{2502.02737}~[cs.CL]
\urldef\tempurl%
\url{https://arxiv.org/abs/2502.02737}
\showURL{%
\tempurl}


\bibitem[Behroozi et~al\mbox{.}(2019)]%
        {behroozi2019}
\bibfield{author}{\bibinfo{person}{Mahnaz Behroozi}, \bibinfo{person}{Chris Parnin}, {and} \bibinfo{person}{Titus Barik}.} \bibinfo{year}{2019}\natexlab{}.
\newblock \showarticletitle{Hiring is Broken: What Do Developers Say About Technical Interviews?}. In \bibinfo{booktitle}{\emph{2019 IEEE Symposium on Visual Languages and Human-Centric Computing (VL/HCC)}}. \bibinfo{pages}{1--9}.
\newblock
\href{https://doi.org/10.1109/VLHCC.2019.8818836}{doi:\nolinkurl{10.1109/VLHCC.2019.8818836}}


\bibitem[Behroozi et~al\mbox{.}(2020a)]%
        {behroozi2020}
\bibfield{author}{\bibinfo{person}{Mahnaz Behroozi}, \bibinfo{person}{Shivani Shirolkar}, \bibinfo{person}{Titus Barik}, {and} \bibinfo{person}{Chris Parnin}.} \bibinfo{year}{2020}\natexlab{a}.
\newblock \showarticletitle{Debugging hiring: what went right and what went wrong in the technical interview process}. In \bibinfo{booktitle}{\emph{Proceedings of the ACM/IEEE 42nd International Conference on Software Engineering: Software Engineering in Society}} (Seoul, South Korea) \emph{(\bibinfo{series}{ICSE-SEIS '20})}. \bibinfo{publisher}{Association for Computing Machinery}, \bibinfo{address}{New York, NY, USA}, \bibinfo{pages}{71–80}.
\newblock
\showISBNx{9781450371254}
\href{https://doi.org/10.1145/3377815.3381372}{doi:\nolinkurl{10.1145/3377815.3381372}}


\bibitem[Behroozi et~al\mbox{.}(2020b)]%
        {behrooziDoes2020}
\bibfield{author}{\bibinfo{person}{Mahnaz Behroozi}, \bibinfo{person}{Shivani Shirolkar}, \bibinfo{person}{Titus Barik}, {and} \bibinfo{person}{Chris Parnin}.} \bibinfo{year}{2020}\natexlab{b}.
\newblock \showarticletitle{Does stress impact technical interview performance?}. In \bibinfo{booktitle}{\emph{Proceedings of the 28th ACM Joint Meeting on European Software Engineering Conference and Symposium on the Foundations of Software Engineering}} (Virtual Event, USA) \emph{(\bibinfo{series}{ESEC/FSE 2020})}. \bibinfo{publisher}{Association for Computing Machinery}, \bibinfo{address}{New York, NY, USA}, \bibinfo{pages}{481–492}.
\newblock
\showISBNx{9781450370431}
\href{https://doi.org/10.1145/3368089.3409712}{doi:\nolinkurl{10.1145/3368089.3409712}}


\bibitem[Chou et~al\mbox{.}(2022)]%
        {Chou2022}
\bibfield{author}{\bibinfo{person}{Yi-Chi Chou}, \bibinfo{person}{Felicia~R. Wongso}, \bibinfo{person}{Chun-Yen Chao}, {and} \bibinfo{person}{Han-Yen Yu}.} \bibinfo{year}{2022}\natexlab{}.
\newblock \showarticletitle{An AI Mock-interview Platform for Interview Performance Analysis}. In \bibinfo{booktitle}{\emph{2022 10th International Conference on Information and Education Technology (ICIET)}}. \bibinfo{pages}{37--41}.
\newblock
\href{https://doi.org/10.1109/ICIET55102.2022.9778999}{doi:\nolinkurl{10.1109/ICIET55102.2022.9778999}}


\bibitem[Cui et~al\mbox{.}(2024)]%
        {Cui2024}
\bibfield{author}{\bibinfo{person}{Jialin Cui}, \bibinfo{person}{Runqiu Zhang}, \bibinfo{person}{Fangtong Zhou}, \bibinfo{person}{Ruochi Li}, \bibinfo{person}{Yang Song}, {and} \bibinfo{person}{Ed Gehringer}.} \bibinfo{year}{2024}\natexlab{}.
\newblock \showarticletitle{How Much Effort Do You Need to Expend on a Technical Interview? A Study of LeetCode Problem Solving Statistics}. In \bibinfo{booktitle}{\emph{2024 36th International Conference on Software Engineering Education and Training (CSEE\&T)}}. \bibinfo{pages}{1--10}.
\newblock
\href{https://doi.org/10.1109/CSEET62301.2024.10663022}{doi:\nolinkurl{10.1109/CSEET62301.2024.10663022}}


\bibitem[Ford et~al\mbox{.}(2017)]%
        {Ford2017}
\bibfield{author}{\bibinfo{person}{Denae Ford}, \bibinfo{person}{Titus Barik}, \bibinfo{person}{Leslie Rand-Pickett}, {and} \bibinfo{person}{Chris Parnin}.} \bibinfo{year}{2017}\natexlab{}.
\newblock \showarticletitle{The Tech-Talk Balance: What Technical Interviewers Expect from Technical Candidates}. In \bibinfo{booktitle}{\emph{2017 IEEE/ACM 10th International Workshop on Cooperative and Human Aspects of Software Engineering (CHASE)}}. \bibinfo{pages}{43--48}.
\newblock
\href{https://doi.org/10.1109/CHASE.2017.8}{doi:\nolinkurl{10.1109/CHASE.2017.8}}


\bibitem[Hall~Jr. and Gosha(2018)]%
        {halljrEffects2018}
\bibfield{author}{\bibinfo{person}{Phillip Hall~Jr.} {and} \bibinfo{person}{Kinnis Gosha}.} \bibinfo{year}{2018}\natexlab{}.
\newblock \showarticletitle{The {{Effects}} of {{Anxiety}} and {{Preparation}} on {{Performance}} in {{Technical Interviews}} for {{HBCU Computer Science Majors}}}. In \bibinfo{booktitle}{\emph{Proceedings of the 2018 {{ACM SIGMIS Conference}} on {{Computers}} and {{People Research}}}}. \bibinfo{publisher}{ACM}, \bibinfo{address}{Buffalo-Niagara Falls NY USA}, \bibinfo{pages}{64--69}.
\newblock
\showISBNx{978-1-4503-5768-5}
\href{https://doi.org/10.1145/3209626.3209707}{doi:\nolinkurl{10.1145/3209626.3209707}}


\bibitem[Kapoor and Gardner-McCune(2020)]%
        {kapoor2020}
\bibfield{author}{\bibinfo{person}{Amanpreet Kapoor} {and} \bibinfo{person}{Christina Gardner-McCune}.} \bibinfo{year}{2020}\natexlab{}.
\newblock \showarticletitle{Barriers to Securing Industry Internships in Computing}. In \bibinfo{booktitle}{\emph{Proceedings of the Twenty-Second Australasian Computing Education Conference}} (Melbourne, VIC, Australia) \emph{(\bibinfo{series}{ACE'20})}. \bibinfo{publisher}{Association for Computing Machinery}, \bibinfo{address}{New York, NY, USA}, \bibinfo{pages}{142–151}.
\newblock
\showISBNx{9781450376860}
\href{https://doi.org/10.1145/3373165.3373181}{doi:\nolinkurl{10.1145/3373165.3373181}}


\bibitem[Lunn et~al\mbox{.}(2021)]%
        {lunn2021}
\bibfield{author}{\bibinfo{person}{Stephanie Lunn}, \bibinfo{person}{Monique Ross}, \bibinfo{person}{Zahra Hazari}, \bibinfo{person}{Mark~Allen Weiss}, \bibinfo{person}{Michael Georgiopoulos}, {and} \bibinfo{person}{Kenneth Christensen}.} \bibinfo{year}{2021}\natexlab{}.
\newblock \showarticletitle{The Impact of Technical Interviews, and other Professional and Cultural Experiences on Students' Computing Identity}. In \bibinfo{booktitle}{\emph{Proceedings of the 26th ACM Conference on Innovation and Technology in Computer Science Education V. 1}} (Virtual Event, Germany) \emph{(\bibinfo{series}{ITiCSE '21})}. \bibinfo{publisher}{Association for Computing Machinery}, \bibinfo{address}{New York, NY, USA}, \bibinfo{pages}{415–421}.
\newblock
\showISBNx{9781450382144}
\href{https://doi.org/10.1145/3430665.3456362}{doi:\nolinkurl{10.1145/3430665.3456362}}


\bibitem[Lunn(2021)]%
        {lunnUneven2021}
\bibfield{author}{\bibinfo{person}{Stephanie~J Lunn}.} \bibinfo{year}{2021}\natexlab{}.
\newblock \showarticletitle{Uneven playing field: Examining preparation for technical interviews in computing and the role of cultural experiences}. In \bibinfo{booktitle}{\emph{2021 ASEE Virtual Annual Conference Content Access}}.
\newblock


\bibitem[McDonald et~al\mbox{.}(2019)]%
        {mcdonald_reliability_2019}
\bibfield{author}{\bibinfo{person}{Nora McDonald}, \bibinfo{person}{Sarita Schoenebeck}, {and} \bibinfo{person}{Andrea Forte}.} \bibinfo{year}{2019}\natexlab{}.
\newblock \showarticletitle{Reliability and {Inter}-rater {Reliability} in {Qualitative} {Research}: {Norms} and {Guidelines} for {CSCW} and {HCI} {Practice}}.
\newblock \bibinfo{journal}{\emph{Proceedings of the ACM on Human-Computer Interaction}} \bibinfo{volume}{3}, \bibinfo{number}{CSCW} (\bibinfo{date}{Nov.} \bibinfo{year}{2019}), \bibinfo{pages}{1--23}.
\newblock
\showISSN{2573-0142}
\href{https://doi.org/10.1145/3359174}{doi:\nolinkurl{10.1145/3359174}}


\bibitem[OpenAI(2025)]%
        {openai2025chatgpt}
\bibfield{author}{\bibinfo{person}{OpenAI}.} \bibinfo{year}{2025}\natexlab{}.
\newblock \bibinfo{title}{ChatGPT-4}.
\newblock
\urldef\tempurl%
\url{https://openai.com/chatgpt}
\showURL{%
\tempurl}
\newblock
\shownote{Large language model}.


\bibitem[Salvi et~al\mbox{.}(2017)]%
        {Salvi2017}
\bibfield{author}{\bibinfo{person}{Vikash Salvi}, \bibinfo{person}{Adnan Vasanwalla}, \bibinfo{person}{Niriksha Aute}, {and} \bibinfo{person}{Abhijit Joshi}.} \bibinfo{year}{2017}\natexlab{}.
\newblock \showarticletitle{Virtual Simulation of Technical Interviews}. In \bibinfo{booktitle}{\emph{2017 International Conference on Computing, Communication, Control and Automation (ICCUBEA)}}. \bibinfo{pages}{1--6}.
\newblock
\href{https://doi.org/10.1109/ICCUBEA.2017.8463876}{doi:\nolinkurl{10.1109/ICCUBEA.2017.8463876}}


\bibitem[Stivers et~al\mbox{.}(2009)]%
        {Stivers2009}
\bibfield{author}{\bibinfo{person}{Tanya Stivers}, \bibinfo{person}{N.~J. Enfield}, \bibinfo{person}{Penelope Brown}, \bibinfo{person}{Christina Englert}, \bibinfo{person}{Makoto Hayashi}, \bibinfo{person}{Trine Heinemann}, \bibinfo{person}{Gertie Hoymann}, \bibinfo{person}{Federico Rossano}, \bibinfo{person}{Jan~Peter de Ruiter}, \bibinfo{person}{Kyung-Eun Yoon}, {and} \bibinfo{person}{Stephen~C. Levinson}.} \bibinfo{year}{2009}\natexlab{}.
\newblock \showarticletitle{Universals and cultural variation in turn-taking in conversation}.
\newblock \bibinfo{journal}{\emph{Proceedings of the National Academy of Sciences}} \bibinfo{volume}{106}, \bibinfo{number}{26} (\bibinfo{year}{2009}), \bibinfo{pages}{10587--10592}.
\newblock
\href{https://doi.org/10.1073/pnas.0903616106}{doi:\nolinkurl{10.1073/pnas.0903616106}}
\showeprint{https://www.pnas.org/doi/pdf/10.1073/pnas.0903616106}


\bibitem[Team(2021)]%
        {SileroVAD}
\bibfield{author}{\bibinfo{person}{Silero Team}.} \bibinfo{year}{2021}\natexlab{}.
\newblock \bibinfo{title}{Silero VAD: pre-trained enterprise-grade Voice Activity Detector (VAD), Number Detector and Language Classifier}.
\newblock \bibinfo{howpublished}{\url{https://github.com/snakers4/silero-vad}}.
\newblock


\end{thebibliography}


\end{document}